\RequirePackage{letltxmacro}
\LetLtxMacro{\LaTeXtextbf}{\textbf}

\documentclass{ieeeaccess}

\LetLtxMacro{\textbf}{\LaTeXtextbf}
\usepackage{cite}
\usepackage{amsmath,amssymb,amsfonts}
\usepackage{algorithmic}
\usepackage{soul}

\usepackage{hyperref}
\usepackage{graphicx}
\usepackage{textcomp}
\usepackage{subcaption}
\usepackage{url}
\usepackage{wrapfig, framed, caption}
\usepackage[ruled, lined, linesnumbered, commentsnumbered, longend]{algorithm2e}
\usepackage{xcolor}

\usepackage{multirow}
\usepackage{enumitem}
\usepackage{booktabs}
\usepackage{array}
\usepackage{float}

\def\BibTeX{{\rm B\kern-.05em{\sc i\kern-.025em b}\kern-.08em
    T\kern-.1667em\lower.7ex\hbox{E}\kern-.125emX}}
\begin{document}
\history{Date of publication xxxx 00, 0000, date of current version xxxx 00, 0000.}
\doi{10.1109/ACCESS.2017.DOI}

\title{A Generic Performance Model for Deep Learning in a Distributed Environment}
\author{\uppercase{Tulasi Kavarakuntla}\authorrefmark{1}, 
\uppercase{Liangxiu Han\authorrefmark{1}*, Huw Lloyd, MIEEE\authorrefmark{1}, Annabel Latham, SMIEEE\authorrefmark{1},  Anthony Kleerekoper\authorrefmark{1}, and Samson B. Akintoye\authorrefmark{1} }
\address[1]{Department of Computing and Mathematics, Manchester Metropolitan University, UK (e-mail: Tulasi.Kavarakuntla@stu.mmu.ac.uk;  l.han@mmu.ac.uk; Huw.Lloyd@mmu.ac.uk; A.Latham@mmu.ac.uk; Akleerekoper@mmu.ac.uk; s.akintoye@mmu.ac.uk )}
}

\markboth
{Author \headeretal: Preparation of Papers for IEEE TRANSACTIONS and JOURNALS}
{Author \headeretal: Preparation of Papers for IEEE TRANSACTIONS and JOURNALS}

\corresp{* Corresponding author: L. Han (e-mail: l.han@mmu.ac.uk).}

\begin{abstract}
Performance modelling of a deep learning application is essential to improve and quantify the efficiency of the model framework. However, existing performance models are mostly case-specific, with limited capability for the new deep learning frameworks/applications. In this paper, we propose a generic performance model of an application in a distributed environment with a generic expression of the application execution time that considers the influence of both intrinsic factors/operations (e.g. algorithmic parameters/internal operations) and extrinsic scaling factors (e.g. the number of processors, data chunks and batch size). We formulate it as a global optimization problem and solve it using regularization on a cost function and differential evolution algorithm to find the best-fit values of the constants in the generic expression to match the experimentally determined computation time. We have evaluated the proposed model on three deep learning frameworks (i.e., TensorFlow, MXnet, and Pytorch). The experimental results show that the proposed model can provide accurate performance predictions and interpretability. In addition, the proposed work can be applied to any distributed deep neural network without instrumenting the code and provides insight into the factors affecting performance and scalability.
\end{abstract}


\titlepgskip=-15pt

\maketitle

\section{Introduction}\label{sec:introduction}
Deep neural networks are effective tools for unsupervised data exploration to discover correlation structures. As a result, they are widely used in computer vision, self-driving cars, medical image analysis, video games, and online self-service applications. However, deep neural network architectures such as GoogLeNet\cite{ballester2016performance}, ResNet\cite{targ2016resnet}, VGG Net\cite{wang2015places205}, and Deep  Convolutional Neural Networks (CNN) \cite{aloysius2017review} necessitate the use of high computational resources. Training with a large amount of data requires a parallelised and distributed environment, primarily data parallelism, model parallelism, pipeline parallelism, and hybrid parallelism. Performance modelling is essential in quantifying the efficiency of large parallel workloads. Performance models are used to obtain run-time estimates by modelling various aspects of an application on a target system. However, accurate performance modelling is a challenging task. Existing performance models are broadly categorised into two catergories: analytical modelling and empirical modelling. Analytical modelling uses a transparent approach to convert a model's or an application's internal mechanism into a mathematical model corresponding to the system's goals, which can significantly expedite the creation of a performance model for the intended system. The existing analytical modelling works investigated deep learning performance modelling and scaling optimization in distributed environments \cite{yan2015performance}, asynchronous GPU processing based on mini-batch SGD\cite{oyama2016predicting}, efficient GPU utilisation in deep learning\cite{song2017towards}, comprehensive analysis and comparison of the performance of deep learning frameworks running on GPUs \cite{shi2016benchmarking, shi2018performance}. However, the major limitation of these works is the poor presentation of the underlying internal operations (i.e., areas of the features' space or specific workload conditions) in the distributed environment.\\
Empirical modelling is a good alternative to analytical models, which predicts the outcome of an unknown set of system parameters based on observation and experimentation. It characterises an algorithm's performance across problem instances and parameter configurations based on sample data. Existing works investigated deep convolutional neural networks using asynchronous stochastic gradient descent techniques in a distributed environment\cite{oyama2016predicting}. Nevertheless, the existing empirical modelling methods are still facing a challenge on how to provide an unbiased experimental study in a distributed environment using GPUs. 


Thus, inspired by the hybridisation of the analytical and empirical approaches, this paper proposes a novel generic performance model that provides a general expression of intrinsic and extrinsic factors of a deep neural network framework in a distributed environment for accurate performance prediction and interpretability.  Especially, the proposed model is applicable to any to any distributed deep neural network without instrumenting the code, which furthermore allows for explaining intrinsic parameters' performance and scalability, providing added value in the field. Our contributions include the following:

\begin{itemize}
\item We have developed a generic expression for a performance model considering the influence of intrinsic parameters and extrinsic scaling factors that affect computing time in a distributed environment. 
\item We have formulated the generic expression as a global optimization problem using regularization on a cost function in terms of the unknown constants in the generic expression, which we have solved using differential evolution to find the best fitting values to match experimentally determined computation times.
\item  We have evaluated the proposed model in three deep learning frameworks, i.e., TensorFlow, MXnet, and Pytorch, to demonstrate its performance efficiency. 
\end{itemize}

The remainder of the paper is organised as follows.
Section \ref{rel} discusses related work of the existing performance models in a distributed environment. In section \ref{Meth}, we discuss research methodology, e.g., problem description, the proposed performance model. Section \ref{exp} discusses an experimental evaluation of the effectiveness of our proposals. Finally, section \ref{con} concludes the paper and highlights the future works.

\section{Related Works}\label{rel}
This section provides an overview of the existing performance models in a distributed computing environment and differential evolution as a solution to the optimization problem.

\subsection{Existing Performance Models}
Performance modelling involves prediction of the performance of a system, the impact of change on an existing system, or the impact of a change in workload on an existing system \cite{4606673, fahringer2004teuta}. Existing performance modelling of Deep Learning (DL) frameworks can be broadly divided into two categories: 1) Analytical modelling and 2) Empirical modelling.

\subsubsection{Analytical Performance Modelling of DL frameworks}
Yan {\em et al.} \cite{yan2015performance} developed a performance model to evaluate the effect of the partitioning and resourcing decisions on the distributed system architectures' overall performance and scalability using a DL framework Adam\cite{chilimbi2014project}.  In addition, the performance model was also used to guide the development of a scalability optimizer that quickly selects the optimal system configuration for reducing DNN training time. However, the model can only be applied to specific DL systems, particularly when it has parameter servers and synchronous weights between worker nodes dynamically.

Qi {\em et al.} \cite{qi2016paleo} developed an analytical performance model named Paleo, predicting the deep neural network performance by considering communication schemes, network architecture and parallelization strategies. The results demonstrated that hybrid parallelism performed much better than data parallelism while training the Alexnet model. However, the model did not consider other factors affecting the overall performance of a model, such as memory usage, data transfer, or communication overhead in distributed environments.

Heehoon Kim {\em et al.} \cite{kim2017performance} evaluated five popular deep learning frameworks TensorFlow\cite{abadi2016tensorflow}, CNTK\cite{seide2016cntk}, Theano\cite{al2016theano}, Caffe-MPI\cite{awan2017s} and Torch\cite{collobert2002torch} in terms of their performance in both single and multi-GPU environments. In this work, each framework incorporated and compared different convolution algorithms, such as Winograd, General Matrix Multiplication (GEMM),  Fast Fourier Transformation (FFT), and direct convolution algorithms, in terms of layered-wise analysis and execution time. The results have shown that FFT and Winograd algorithms surpass the GEMM and other convolution algorithms. However, the convolution algorithms used by the frameworks provided poor explainability regarding their internal operations.

Shi {\em et al.} \cite{shi2018performance} proposed a performance model to evaluate various distributed deep learning frameworks' performance with different convolutional neural networks in the multi-GPU environment. They measured training time, memory usage, and GPU utilization and compared the frameworks in terms of training time and resource utilization. However, they did not provide a breakdown of the time to divide the mini-batch into smaller batches or measure the load imbalance factor, which are critical factors that could significantly affect the training efficiency and performance in a parallel computing environment. Kavarakuntla {\em et al.} \cite{multigpuperformance} extended Shi's analytical performance model to evaluate the deep learning frameworks' run-time performance with the autoencoder, multilayer perceptron and convolutional neural network models in the GPU cluster environment. The extended model considered the load imbalance factor and made a layer-wise analysis of a neural network, providing a more comprehensive evaluation of the frameworks' performance. The experimental results showed that the load balance is an influential factors affecting the system performance. 

However, the models mentioned above have poor explainability and were developed for specific architectures, which were not generic and couldn't be applied to a wide range of neural networks.

\subsubsection{Empirical Modelling of DL frameworks}
Empirical modelling builds models through observation and experimentation, which is antithetical to analytical modelling.

Oyama {\em et al.} \cite{oyama2016predicting} proposed a performance model for predicting the statistics of an asynchronous stochastic gradient descent-based deep learning system, potentially improving the model's performance by optimizing the hyperparameters, such as gradient staleness and mini-batch size. They did not consider parallelization methods and applied direct weights synchronized among GPUs. The study results showed that the proposed method could predict the statistics of asynchronous SGD parameters, including mini-batch size, sweeping dataset time, staleness, and probability distributions of these essential parameters. However, the work did not address the issue of communication overhead and network latency, which could significantly affect the performance of distributed deep learning systems.

Rakshith {\em et al.} \cite{rakshith2022performance} presented an empirical study of the performance of Horovod, a distributed deep learning framework, for image classification tasks. They evaluated the performance of Horovod on two popular image datasets, CIFAR-10 and ImageNet, using a cluster of machines with varying numbers of GPUs. They also compared the performance of Horovod with other distributed deep learning frameworks, such as PyTorch and TensorFlow, and found that Horovod achieved better performance in certain scenarios. They provided recommendations for optimizing the performance of Horovod on large-scale image datasets, such as using efficient data loading and preprocessing techniques, and optimizing the communication and synchronization between the machines. However, the experimental configuration utilized in the research might not accurately reflect real-world situations in which the underlying hardware and network setups may differ substantially.

Lin {\em et al.} \cite{lin2020topology} considered the network topology and communication patterns to train deep learning models on GPU clusters. The model included the communication and computation times for each layer in a deep neural network and used a prediction model that is more sophisticated than a simple linear regression approach to predict the total training time. The model was evaluated on several deep learning benchmarks and showed that it achieved higher accuracy in predicting training time than existing models. The model could also be used to optimize the performance of distributed deep learning by finding the optimal configuration of GPU nodes and reducing the training time. However, the assessment of the proposed model was confined to three distinct GPU clusters, potentially limiting its generalizability to other GPU clusters or distributed DL architectures.

Most recently, inspired by the concept of combining elements of analytical modelling and empirical modelling for better performance prediction developed in other fields \cite{didona2015enhancing}, we \cite{performancemodel} developed a generic performance model for deep learning applications in distributed environments, which offers the advantage of applicability to various deep learning frameworks. However, its performance is sub-optimal and lacks comprehensive analysis and experimental evaluation.

To address the above limitations, in this paper, we have implemented the model that gives insights into the intrinsic parameters' performance and scalability of the extrinsic parameters for more accurate performance prediction. Our proposed model in this paper provides a generic expression applicable to any distributed deep neural network without instrumenting the code and enabling functionality such as explaining internal parameters' performance and scalability.  The detailed method is described in section III below.

\subsection{ Differential Evolution}
Differential Evolution (DE) was developed by Storn et al.\cite{de:storn} as an algorithm to solve various complex optimization problems such as motor fault diagnosis \cite{sym13071291}, structure prediction of materials \cite{Yang2021CrystalSP}, automatic clustering techniques \cite{de:Saha}, community detection \cite{sym9090183}, learning applications \cite{de:Marco} and so on. 
The algorithm achieves optimal solutions by maintaining a population of individual solutions and employing a distinct process to generate new offspring through the combination of existing solutions. Those offspring exhibiting superior objective values are retained in subsequent iterations of the algorithm, thereby enhancing the individual's new objective value and subsequently assimilating them into the population. 
Conversely, if the newly acquired objective value fails to surpass existing solutions, it is promptly disregarded. This iterative process continues until a specific termination condition is met, ensuring the algorithm's convergence  \cite{AHMAD20223831}. It shares similarities with other evolutionary algorithms, such as the Genetic Algorithm (GA) \cite{de:Liu}, wherein mutation, crossover, and selection operators are employed to steer the population towards increasingly favourable solutions. In contrast to the genetic algorithm, the differential evolution algorithm imparts mutation to each individual while transferring them to the next generation. In its mutation procedure, for each solution, three more individuals are picked from the population, and as a consequence, a mutated individual is produced. It is determined based on the fitness value whether or not the first individual selected will be replaced. In differential evolution, the crossover is not the primary operation, as it is in the genetic algorithm. In recent times, several works have been proposed to use DE for neural network optimization \cite{ math8010069}, \cite{de:Ikushima}, \cite{9504878}. 

However, none of the works mentioned above used DE to analyse and evaluate the performance of deep neural networks with many processes in a distributed environment with the goal of finding the best-fit values by minimising the regularised cost function.


\section{Methodology}\label{Meth}
\subsection{The Generic Performance Model}
Given an application consisting of a number of processes in a distributed environment, the execution time of the application can be considered from two levels: 1) Execution time of internal processes of the application (for example, intrinsic parameters of the application); and 2) External scaling factors that affect the computing efficiency (such as a number of machines/processors or data chunks or batch size). A generic performance model for computing total computational time $(t)$ per iteration of an application can be described as follows:
\begin{align}
\label{eq:1}
 t(I,E) = t_{I}(I) f_{E}(E) + C
\end{align}
Here, we represent intrinsic parameters as $I$, $E$ represents the extrinsic parameters, $t_{I}$ represents the computation time of the processes affected by intrinsic parameters, $f_{E}$ represents extrinsic scaling factors that affect the computing performance, and $C$ is a constant. In general, $I$ and $E$ are vectors in which each element is a hyperparameter of the deep learning model such as a filter size (intrinsic) or batch size (extrinsic). 

\begin{figure*}[h!]
 \centering
\includegraphics[width=0.9\textwidth]{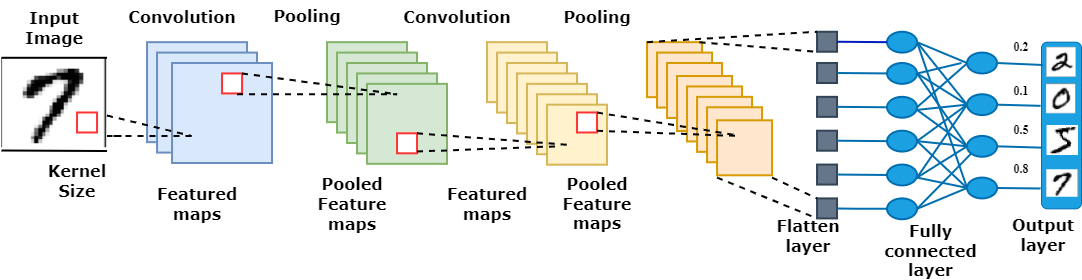}
\caption{Internal processes involved in a convolutional neural network.} \label{Fig:1}
\end{figure*}

\begin{figure}[!t]
\centering
\includegraphics[width=3.5in]{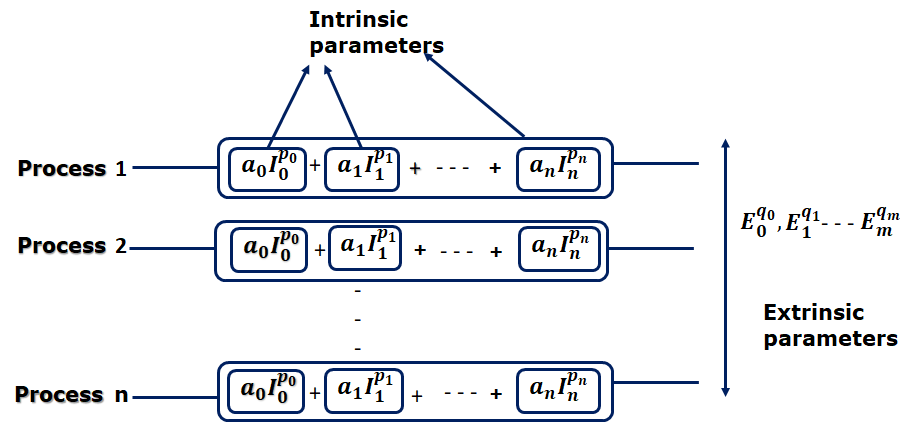} 
\caption{Functional diagram of proposed performance model.}\label{Fig:2}
\label{fig:Performance model}
\end{figure}

In our model, we represent the internal time $t_{I}$ as a sum of terms in powers of the components of $I$:
\begin{align}
\label{eq:2}
     t_{I} = \sum_{i=1}^{n} a_{i}I_{i}^{p_{i}}
\end{align}
Basically, intrinsic parameters represent model parameters of the deep neural network, as shown in Fig.\ref{Fig:1}. In equation \eqref{eq:2}, the coefficients $a_{i}$ relate to the relative importance of the processes, and the powers $p_{i}$ relate to the computational complexity. 
The external factors are related to scaling, and these appear in the model as multiplicative terms with different powers in the computation of the external scaling factor $f_{E}$, which is given by:
\begin{align}
\label{eq:3}
  f_E=\prod_{j=1}^{m}E_{j}^{q_{j}}
\end{align}
Here, the powers $q_{j}$ give information about scalability.
By substituting the $t_{I}$ and $f_{E}$ in equation \eqref{eq:1}, the computational time (t) is given as follows:
\begin{align}
\label{eq:4}
t(I,E,x) = \left ( \sum_{i=1}^{n} a_{i}I_{i}^{p_{i}}\right)\prod_{j=1}^{m}E_{j}^{q_{j}}+C
\end{align}
which we now write as a function of $I$, $E$ and $x$ where $\boldsymbol{x}=\left \{ {a_{1},...,\,\,a_{n_{I}}, p_{1},...\,p_{n_{I}}, q_{1},...\,q_{n_{E}}, c} \right \} \in \mathbb{R}^M$ is a vector formed by combining $\boldsymbol{a}, \boldsymbol{p}, \boldsymbol{q}$ and the constant coefficient $\boldsymbol{C}$. In \eqref{eq:4}, the intrinsic parameters $I$ and extrinsic parameters $E$ are the known input values. $\boldsymbol{a}, \boldsymbol{p}, \boldsymbol{q}$ and coefficient $\boldsymbol{C}$ are unknown constants. 
The functional diagram of the proposed performance model is shown in Fig.~\ref{Fig:2}.

We compute the optimal values of these unknown constants (total: $M = 2n_{I}+n_{E}+1$) using the differential evolution algorithm. The aim is to find the best-fitting values of these constants, by fitting the model to experimentally measured execution times obtained with different values of the internal and external parameters $I$ and $E$. Before going to the cost function formulation of the differential evolution algorithm \cite{fleetwood2004introduction}, we describe the general methodology for obtaining the experimental data. For every possible combination of values of intrinsic and extrinsic parameter there will, in general, be too many combinations for an exhaustive grid search. Therefore, we have applied random sampling to ensure that every hyperparameter in the population has an equal opportunity of being selected for obtaining measured times. Here, the methodology used for measured time is the time taken for an iteration of an epoch. We compute the iteration time as the difference between an iteration's end time and starting time.

The experimental data for fitting the model comprises $N$ measurements with randomly selected values of the parameters. We denote the values of the intrinsic parameters by
\begin{equation}
I_{i,k},\hspace{5pt} \hspace{5pt} i\in [1,n_{I}],\hspace{5pt}k\in [1,N]
\end{equation}
where $i$ indexes the components of the vector of parameters, and $k$ refers to a given observation in the experiment. Similarly, the extrinsic parameters are denoted by
\begin{equation}
E_{j,k},\hspace{5pt} \hspace{5pt} j\in [1,n_{E}],\hspace{5pt}k\in [1,N]
\end{equation}
The measured time for obervation $k$ is
\begin{equation}
t_{k} ,\hspace{5pt} k \in [1,N].
\end{equation} 
Here, $i,j$ denote the input feature indices. $k \in [1,N]$ indicate the sample index in dataset $\boldsymbol{D}$. $N$ is the number of input samples in $\boldsymbol{D}$.

\subsection{Global Optimization Using Differential Evolution}
Given the generic expression as shown in equation \eqref{eq:4}, as mentioned in earlier sub section, we find the best fit values of $\boldsymbol{a}, \boldsymbol{p}, \boldsymbol{q}$ and $C$ by minimizing a cost function. We formulate the cost function as the mean absolute difference between the predicted execution time and the actual measured times as follows:

\begin{equation} \label{eq:cost}
f(x) = \frac{1}{N}\sum_{k=1}^{N} \left | t_{k}-\hat{t}_{k} (I_k, E_k, x) \right |
\end{equation}
where $N$ as number of data samples, $t_{k}$ is the measured time, $\hat{t}_{k}$ is the predicted time derived from equation~\ref{eq:4}.

To solve the above optimization problem, we have used the differential evolution algorithm (DE) using the cost function in equation~\ref{eq:4}. The mean absolute error between the predicted times of the model and the measured times is minimized, resulting in a value of the vector $x$ which represents the best-fitting model. Recall that this vector encodes the coefficients and powers of the terms in the model due to each of the hyperparamters; these can then be used to make predictions of the execution time for any set of values of the intrinsic and extrinsic paramters and furthermore, inspection of these coefficients can provide insight into the relative importance and computational complexity of the internal processes, as well as the scalability of the external processes. 
For this work, we use the DE implementation from the {\em scipy} python package, with default values of the hyperparameters. We enforce limits of $(0 \ldots 1000)$ for constants and coefficients ($a, C$) and $-5 \ldots 5$ for powers ($p, q$).

\subsection{Regularization}
A globally optimized, unconstrained model may be prone to overfitting or producing unstable solutions with high parameter variance. To address these issues, we introduce a regularization term to the cost function. Regularization achieves the best fit by introducing a penalizing term in the cost function, which assigns a higher penalty to complex curves. So, we are motivated to apply regularization to our performance model. Generally, regularization can be defined as:

\begin{equation}
f_{\mbox{\scriptsize reg}}(x) = f(x) + \lambda .L
\end{equation}

The parameter  $\lambda$ controls the balance between bias and variance in the model, where L represents the model's complexity. There are two regularization techniques: (a) Lasso regression (L1) and (b) Ridge regression (L2). L1 regularization, also known as lasso regression, includes the absolute value of the coefficient magnitude as a penalty in the loss function. The resulting solution from L1 regularization is sparse, meaning it tends to eliminate less important features by setting their coefficients to zero. This is useful for feature selection when dealing with a large number of features. On the other hand, L2 regularization, or ridge regression, incorporates the squared magnitude of the coefficient as a penalty in the loss function. The solution obtained from L2 regularization is non-sparse and penalizes the model's complexity. 
The regularization parameter $\lambda$ penalizes all parameters except the intercept, ensuring that the model generalizes the data and avoids overfitting. Ridge regression uses the squared magnitude of the coefficient as the penalty term. We have applied both L1 and L2 regularizations to the performance model. The cost function for optimization with both L1 and L2 regularizations is as follows:

\begin{align}
f(x) = \frac{1}{N}\sum_{k=1}^{N} \left | t_{k}-\hat{t}_{k} (I_k, E_k, x) \right | + \lambda .\sum_{k=1}^{N} \left | x \right |  \\
f(x) = \frac{1}{N}\sum_{k=1}^{N} \left | t_{k}-\hat{t}_{k} (I_k, E_k, x) \right | + \lambda .\sum_{k=1}^{N} \left | x \right |^{2}
\end{align}
 Here, applying the regularization term $\lambda$ reduces the bias-variance trade-off in the internal processes.

\section{Experimental Evaluation}\label{exp}
To evaluate the performance of the proposed model, we have applied our approach to three popular deep learning frameworks (TensorFlow, PyTorch and MxNet) and conducted extensive experiments. We assess the proposed performance evaluation approach by modelling distributed training of a CNN architecture on a multi-GPU system. The main goal is to investigate how well the predicted execution time fits the experimentally measured time. 

\subsection{System Configuration}
 We implement the experiments on a single node containing three GEFORCE RTX 2080 GPUs, each with 2.60 GHz speed and 16 GB GPU RAM. The node also consists of a 2.81 GHz speed CPU machine, 25 Gbps network bandwidth and a CUDA-10.2 with a Linux operating system. Furthermore, the node consists of various software configurations/installations, including PyTorch 1.2.0, Torchvision 0.4.0,  Python 3.6, TensorFlow 2.1.0 and MXnet 1.6.0.
\subsection{Dataset and Model Selection}
For three deep learning frameworks, we selected a CNN architecture, LeNet-5, which Yann LeCun proposed in 1998 as a general common neural network structure for handwritten font recognition. It consists of two convolutional layers, two fully-connected layers, pooled layers for cross-combination and an output layer that predicts values via the fully connected layer. Besides, LeNet-5 works well with handwritten datasets \cite{Park:mnist}, it also reduces the number of parameters and can automatically learn features from raw pixels \cite{Khan:lenet}. 

We train LeNet-5 on three popular datasets, MNIST, fashion-MNIST and CIFAR-10, using TensorFlow, PyTorch and MxNet, in a multi-GPU system. 
MNIST \cite{LeCun:mnist, Deng:mnist} is a database of handwritten digits derived by the National Institute of Standards and Technology (NIST) for learning techniques and pattern recognition methods with a little effort on pre-processing and formatting. It contains 60,000 training and 10,000 testing images, divided into four files: training set images, testing set images, training set labels and testing set labels. Each image has 28 x 28 pixels. Fashion-MNIST \cite{Xiao2017FashionMNISTAN} replaces the MNIST, where each image has a 28x28 grayscale and is associated with a label from 10 classes. The CIFAR-10 dataset \cite{Giuste2020CIFAR10IC} comprises 60,000 images, classified into ten classes - aeroplane, automobile, bird, cat, deer, dog, frog, horse, ship, and truck. Each image has $32 \times 32$ pixels, and each classification has 6,000 images.

\subsection{Performance Metrics}
The scalability and the mean absolute percentage error (MAPE) are selected as performance metrics for run-time evaluation on three different frameworks. Scalability is measured in the powers of external parameters as shown in \eqref{eq:3}. The MAPE can be defined as:
\begin{align}
MAPE = \frac{1}{N}\sum_{k=1}^{N}\frac{\left | t_{k}-\hat{t_{k}} \right |}{t_{k}} 
\end{align}
where $t_{k}$ = the measured value, $\hat{t_{k}}$ = the predicted value, n = the total number of data points.

The experimental evaluation aims to evaluate the proposed performance model and find the best-fit values using the differential evolution algorithm. The MAPE is used to evaluate the closeness of this fit and the quality of the {\em performance model}. The scaling parameters are used in our proposed model to evaluate the performance of the deep learning frameworks.

\subsection{Experiments}
We have conducted a set of experiments to evaluate the proposed model from the following aspects: 
\begin{enumerate}

\item Performance evaluation of deep learning frameworks using the proposed performance model with and without regularization. Specifically, we have applied the proposed performance model to three deep learning frameworks:  TensorFlow, MXnet, and PyTorch,  under two circumstances, with and without regularization.

\item Comparison of the proposed model with the existing black-box machine learning models. We have also compared our proposed model with two widely used models including Random Forest Regression \cite{Ho:2019vm} and Support Vector Machine \cite{Cortes:1995wa}, and demonstrated its performance and interpretability.  
\end{enumerate}	

 In our experiment, we have performed the distributed training of LeNet-5 on MNIST, fashion-MNIST and CIFAR-10 datasets using the three deep learning frameworks. The values of the experimental training parameters are created by applying random sampling on a set of intrinsic and extrinsic parameters and its corresponding average training time taken by a deep CNN architecture per iteration. Table \ref{Table-I} shows intrinsic and extrinsic parameters and their possible values. The intrinsic parameters are the model's hyperparameters, including kernel size, pooling size, activation function, etc. The number of GPUs and the batch size are extrinsic factors since these affect the scaling over multiple processes.

\begin{table}[]
\centering
\caption{Parameters of the performance model, with ranges of values sampled in the experiments.}\label{Table-I}
\begin{tabular}{|l|l|l|}
\hline
\bf Index & \bf Name         &\bf Set of possible values considered          \\ \hline
              \multicolumn{3}{|c|}{\bf Intrinsic parameters}                                           \\ \hline
1             & Kernel size          & \{2,3,4,5\}                                \\ \hline
2             & Pooling size         & \{2,3,4,5\}                                \\ \hline
3             & Activation function  & \{Relu, Tanh, Sigmoid\}                    \\ \hline
4             & Optimizer            & \{Adam, SGD\}                              \\ \hline
5             & Image\_dataset\_name & \{MNIST,Fashion-MNIST,CIFAR-10\}            \\ \hline
6             & Number of filters    & \{4,8,16,32,64\}                           \\ \hline
7             & Learning rate        & \{0.1,0.01,0.001,$10^{-4}$, $10^{-5}$, $10^{-6}$\} \\ \hline
8             & Padding\_mode        & \{valid, same\}                            \\ \hline
9             & Stride               & \{1,2,3\}                                  \\ \hline
10            & Dropout probability  & \{0.2,0.5,0.8\}                            \\ \hline
              \multicolumn{3}{|c|}{\bf Extrinsic parameters }                                           \\ \hline
11            & Number of GPUs       & \{1,2,3\}                                  \\ \hline
12            & Batchsize           & \{8,16,32,64,128\}                         \\ \hline
\end{tabular}
\end{table}

The experiments involve several trials in which we measure the time for a single training iteration using randomly selected intrinsic and extrinsic parameter values. We conduct 1500 trials to prepare a dataset of 1500 data samples. For each sample, we run three iterations with the same parameter values, and take the median value of the measured time. The experimental data for 900 trials are used to fit our performance model or train the standard black-box models for comparison. The remaining 600 are used to test and validate models. Finally, the experimental parameters are used to build three performance evaluation models, such as the Differential evolution (DE) algorithm with and without using regularization models and two standard black-box models. We run each fit ten times with different random seeds to obtain the mean and standard deviation for each of our fitted parameters. The performance of these models and their corresponding results are explained in the subsequent subsections.


\begin{table*}[h!]
\begin{center}
\caption{Derived intrinsic and extrinsic parameters from the differential evolution-optimized performance models for the three deep learning frameworks. Parameters are given as the mean and standard deviation over ten fits. $a$ and $p$ represent coefficients and powers, respectively, of a term representing an intrinsic parameter, whereas $q$ is power in a multiplicative term representing an extrinsic (scaling) parameter.}\label{Table-II}

\begin{tabular}{|l|c|c|c|c|c|c|}
\hline
                    & \multicolumn{2}{|c|}{\bf Mxnet} &   \multicolumn{2}{|c|}{\bf Pytorch} &
                    \multicolumn{2}{|c|}{\bf TensorFlow}       
                    \tabularnewline \hline

\bf Intrinsic parameters& $a$                         & $p$     & $a$                           & $p$     & $a$                    & $p$                             \tabularnewline \hline
Filter size         & 554.87 $\pm$ 311.73   & -4.06 $\pm$ 0.53 & 423.36 $\pm$ 256.88   & -2.88 $\pm$ 1.04 & 346.73 $\pm$ 216.24     & -3.22 $\pm$ 0.78 \tabularnewline \hline
Kernel size         & 10.57 $\pm$ 7.05      & -4.10 $\pm$ 0.70 & 168.54 $\pm$ 123.27   & -2.34 $\pm$ 1.82 &  54.78 $\pm$ 32.91      & -4.00 $\pm$ 1.41 \tabularnewline \hline
Pool size           & 18.08 $\pm$ 5.17      & -4.21 $\pm$ 0.46 & 209.14 $\pm$ 186.87   & -3.31 $\pm$ 0.92 & 79.45 $\pm$ 53.53       & -3.48 $\pm$ 1.33 \tabularnewline \hline
Learning rate       & 459.50 $\pm$ 258.52   & 3.68 $\pm$ 0.62  & 489.52 $\pm$ 221.63   & 3.21 $\pm$ 0.70  & 458.34 $\pm$ 278.03     & 3.26 $\pm$ 0.91  \tabularnewline \hline
Stride              & 17.29 $\pm$ 6.12      & -0.83 $\pm$ 0.23 & 140.64 $\pm$ 138.62   & -0.63 $\pm$ 0.58 & 29.00 $\pm$ 14.54       & -1.85 $\pm$ 0.90 \tabularnewline \hline
Dropout 
probability         & 1.79 $\pm$ 0.75       & 2.24 $\pm$ 1.62  & 437.06 $\pm$ 184.32   &  1.80 $\pm$ 1.66 & 10.23 $\pm$ 9.51        & 1.87 $\pm$ 1.62  \tabularnewline \hline
Same                & 2.50 $\pm$ 0.97       &      -       & 11.02 $\pm$ 5.09      &       -      & 6.14 $\pm$ 1.54         &      -       \tabularnewline \hline
Valid               & 1.56 $\pm$ 0.96       &      -       &  0.77 $\pm$ 1.81      &       -      & 1.61 $\pm$ 2.24         &      -       \tabularnewline  \hline
Sigmoid             & 23.25 $\pm$ 10.23     &      -       & 475.92 $\pm$ 139.65   &       -      & 251.57 $\pm$ 122.01     &      -       \tabularnewline \hline
Relu                & 21.90 $\pm$ 10.40     &      -       & 475.56 $\pm$ 137.27   &       -      & 255.93 $\pm$ 122.35     &      -       \tabularnewline \hline
Tanh                & 23.14 $\pm$ 10.30     &      -       & 444.48 $\pm$ 138.25   &       -      & 254.28 $\pm$ 121.27     &      -       \tabularnewline 
\hline
MNIST               & 35.75 $\pm$ 12.81     &      -       & 815.62 $\pm$ 69.44    &       -      & 232.24 $\pm$ 108.77     &      -       \tabularnewline \hline
Fashion-MNIST       & 35.94 $\pm$ 12.75   &      -       & 815.68 $\pm$ 68.39    &       -      & 231.33 $\pm$ 109.93       &      -       \tabularnewline \hline
CIFAR-10            & 18.57 $\pm$ 12.68     &      -       & 308.73 $\pm$ 53.32    &       -      &  124.56 $\pm$ 108.01    &      -       \tabularnewline 
\hline
SGD                 & 16.68 $\pm$ 10.10     &      -       & 361.65 $\pm$ 130.64   &       -      & 158.74 $\pm$ 109.25     &      -       \tabularnewline \hline
Adam                &  16.85 $\pm$ 10.32    &      -       & 720.15 $\pm$ 123.99   &       -      & 168.55 $\pm$ 108.67     &      -       \tabularnewline
\hline

\bf Extrinsic parameters &  \multicolumn{2}{|c|}{$q$} &  \multicolumn{2}{|c|}{$q$} &  \multicolumn{2}{|c|}{$q$}    \tabularnewline \hline
Batchsize           &      \multicolumn{2}{|c|}{-0.99 $\pm$ 0.003} &  \multicolumn{2}{|c|}{-1.13 $\pm$ 0.01} &  \multicolumn{2}{|c|}{-1.35 $\pm$ 0.08}   \tabularnewline \hline
No. of GPUs             &      \multicolumn{2}{|c|}{-0.99 $\pm$ 0.004} &  \multicolumn{2}{|c|}{-1.029 $\pm$ 0.001} &   \multicolumn{2}{|c|}{-0.74 $\pm$ 0.001}  \tabularnewline
\hline

\bf Constant term & \multicolumn{2}{|c|}{$C$} &  \multicolumn{2}{|c|}{$C$} &  \multicolumn{2}{|c|}{$C$}    \tabularnewline \hline
             &      \multicolumn{2}{|c|}{3.703 $\pm$ 0.017} &  \multicolumn{2}{|c|}{12.677 $\pm$ 0.038} &  \multicolumn{2}{|c|}{1.930 $\pm$ 0.122}   \tabularnewline \hline

\end{tabular}
\end{center}
\end{table*}

\begin{table*}[h!]
\begin{center}
\caption{Derived intrinsic and extrinsic parameters from the differential evolution-optimized performance models for the three deep learning frameworks using L2 regularization. Parameters are given as the mean and standard deviation over ten fits. $a$ and $p$ represent coefficients and powers, respectively, of a term representing an intrinsic parameter, whereas $q$ is power in a multiplicative term representing an extrinsic (scaling) parameter.}\label{Table-III}

\begin{tabular}{|l|c|c|c|c|c|c|}
\hline
                    & \multicolumn{2}{|c|}{\bf Mxnet} &   \multicolumn{2}{|c|}{\bf Pytorch} &
                    \multicolumn{2}{|c|}{\bf TensorFlow}       
                    \tabularnewline \hline

\bf Intrinsic parameters& $a$                         & $p$     & $a$                           & $p$     & $a$                    & $p$                             \tabularnewline \hline
Filter size         & 6.27$\pm$ 0.59   & 0.36 $\pm$ 0.01 & 6.07 $\pm$ 1.59   & 0.89 $\pm$ 0.05 & 8.39 $\pm$ 0.37     & 0.77 $\pm$ 0.01 \tabularnewline \hline
Kernel size         & 4.44 $\pm$ 0.65      & 0.50 $\pm$ 0.04 & 4.84 $\pm$ 1.90   & 2.02 $\pm$ 0.24 &  6.59 $\pm$ 0.29      & 2.04 $\pm$ 0.03 \tabularnewline \hline
Pool size           & 4.69 $\pm$ 0.33      & 0.52 $\pm$ 0.03 & 3.23 $\pm$ 0.83   & 1.55 $\pm$ 0.45 & 6.70 $\pm$ 0.67       & 1.98 $\pm$ 0.05 \tabularnewline \hline
Learning rate       & 3.62 $\pm$ 0.41   & -0.04 $\pm$ 0.003  & 3.75 $\pm$ 1.70   & -0.27 $\pm$ 0.02  & 4.40 $\pm$ 0.60     & -0.22 $\pm$ 0.007 \tabularnewline \hline
Stride              & 4.51 $\pm$ 0.40      & -0.99 $\pm$ 0.11 & 2.92 $\pm$ 1.54   & -0.83 $\pm$ 1.42 & 4.13 $\pm$ 0.59       & 2.46 $\pm$ 0.10 \tabularnewline \hline
Dropout 
probability         & 4.20 $\pm$ 0.67       & -0.35 $\pm$ 0.05  & 35.92 $\pm$ 1.15   &  -5.00 $\pm$ 0.00 & 4.46 $\pm$ 0.43        & -1.94 $\pm$ 0.07  \tabularnewline \hline
Same                & 2.66 $\pm$ 0.51       &      -       & 2.08 $\pm$ 0.84      &       -      & 1.90 $\pm$ 0.58         &      -       \tabularnewline \hline
Valid               & 1.50 $\pm$ 0.43       &      -       & -0.57 $\pm$ 1.56      &       -      & 0.49 $\pm$ 0.71         &      -       \tabularnewline  \hline
Sigmoid             & 2.18 $\pm$ 0.45     &      -       & 2.32 $\pm$ 1.15   &       -      & 1.41 $\pm$ 0.41     &      -       \tabularnewline \hline
Relu                & 1.52 $\pm$ 0.33     &      -       & 3.21 $\pm$ 1.35   &       -      & 1.85 $\pm$ 0.64     &      -       \tabularnewline \hline
Tanh                & 2.29 $\pm$ 0.39     &      -       & 2.93 $\pm$ 1.93   &       -      & 1.99 $\pm$ 0.71     &      -       \tabularnewline 
\hline
MNIST               &5.48 $\pm$ 0.52     &      -       & 3.37 $\pm$ 1.35    &       -      & 1.99 $\pm$ 0.72     &      -       \tabularnewline \hline
Fashion-MNIST       & 7.73 $\pm$ 0.36   &      -       & 3.42 $\pm$ 1.56    &       -      & 2.28 $\pm$ 0.72       &      -       \tabularnewline \hline
CIFAR-10            & 1.00 $\pm$ 0.02     &      -       & 1.89 $\pm$ 1.00    &       -      &  1.63 $\pm$ 0.69    &      -       \tabularnewline 
\hline
SGD                 & 2.31 $\pm$ 0.36     &      -       & 2.16 $\pm$ 1.00   &       -      & 1.73 $\pm$ 0.46     &      -       \tabularnewline \hline
Adam                & 1.78 $\pm$ 0.41    &      -       & 3.42 $\pm$ 1.45   &       -      & 2.01 $\pm$ 0.85     &      -       \tabularnewline
\hline

\bf Extrinsic parameters &  \multicolumn{2}{|c|}{$q$} &  \multicolumn{2}{|c|}{$q$} &  \multicolumn{2}{|c|}{$q$}    \tabularnewline \hline
Batchsize           &      \multicolumn{2}{|c|}{-0.87 $\pm$ 0.005} &  \multicolumn{2}{|c|}{-1.00 $\pm$ 0.007} &  \multicolumn{2}{|c|}{-1.19 $\pm$ 0.01}   \tabularnewline \hline
No. of GPUs             &      \multicolumn{2}{|c|}{-1.07 $\pm$ 0.007} &  \multicolumn{2}{|c|}{-1.01 $\pm$ 0.004} &   \multicolumn{2}{|c|}{-0.74 $\pm$ 0.005}  \tabularnewline
\hline

\bf Constant term & \multicolumn{2}{|c|}{$C$} &  \multicolumn{2}{|c|}{$C$} &  \multicolumn{2}{|c|}{$C$}    \tabularnewline \hline
             &      \multicolumn{2}{|c|}{3.45$\pm$ 0.024} &  \multicolumn{2}{|c|}{1.03 $\pm$ 0.07} &  \multicolumn{2}{|c|}{12.62 $\pm$ 0.05}   \tabularnewline \hline

\end{tabular}
\end{center}
\end{table*}

\begin{figure*}[h!]
     \begin{subfigure}[b]{0.3\textwidth}
         \includegraphics[width=\textwidth]{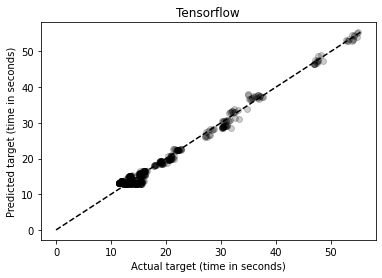}
         \caption{TensorFlow}
     \end{subfigure}
     \hfill
     \begin{subfigure}[b]{0.3\textwidth}
         \includegraphics[width=\textwidth]{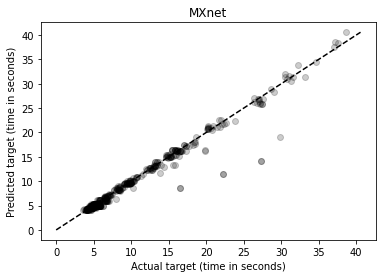}
         \caption{MXnet}
     \end{subfigure}
    \hfill
     \begin{subfigure}[b]{0.3\textwidth}
         \includegraphics[width=\textwidth]{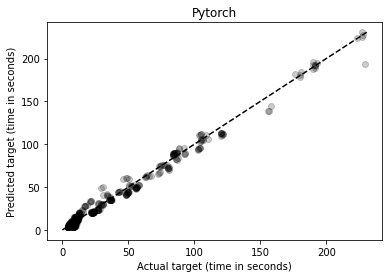}
         \caption{Pytorch}
     \end{subfigure}
     
\caption{The proposed performance model prediction and the actual measured times in three deep learning frameworks without regularization \label{Fig:3}}
    
\end{figure*}

\begin{figure*}[h!]
     \begin{subfigure}[b]{0.3\textwidth}
         \includegraphics[width=\textwidth]{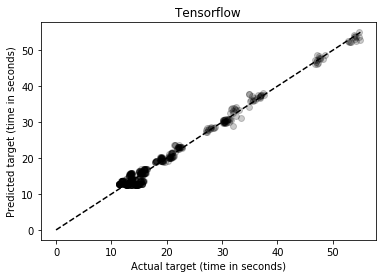}
         \caption{TensorFlow}
     \end{subfigure}
     \hfill
     \begin{subfigure}[b]{0.3\textwidth}
         \includegraphics[width=\textwidth]{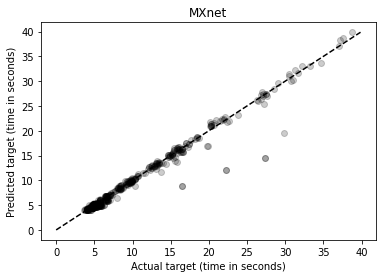}
         \caption{MXnet}
     \end{subfigure}
    \hfill
     \begin{subfigure}[b]{0.3\textwidth}
         \includegraphics[width=\textwidth]{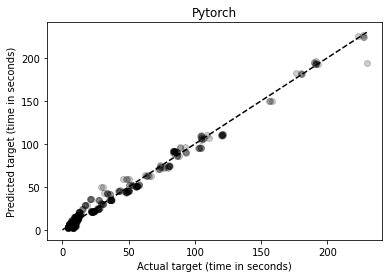}
         \caption{Pytorch}
     \end{subfigure}
     
\caption{The proposed performance model prediction and the actual measured time in three deep learning frameworks with regularization \label{Fig:4}}
    
\end{figure*}

\begin{figure*}[h!]
     \begin{subfigure}[b]{0.3\textwidth}
         \includegraphics[width=\textwidth]{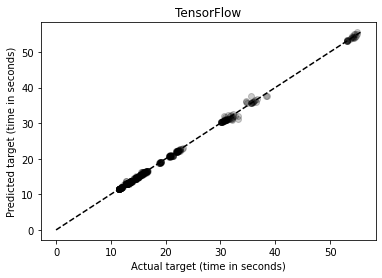}
         \caption{TensorFlow}
     \end{subfigure}
     \hfill
     \begin{subfigure}[b]{0.3\textwidth}
         \includegraphics[width=\textwidth]{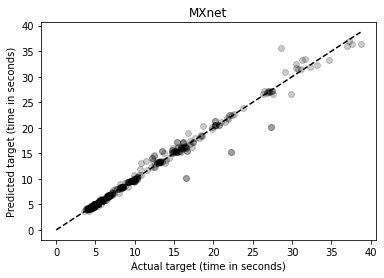}
         \caption{MXnet}
     \end{subfigure}
    \hfill
     \begin{subfigure}[b]{0.3\textwidth}
         \includegraphics[width=\textwidth]{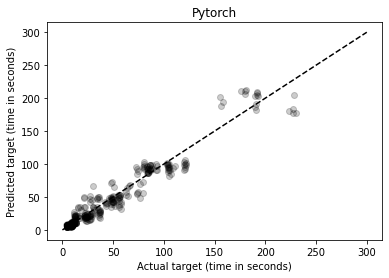}
         \caption{Pytorch}
     \end{subfigure}
     
\caption{Random forest regressor prediction and the actual measured time in three deep learning frameworks \label{fig:5}}
    
\end{figure*}

\begin{figure*}[h!]
     \begin{subfigure}[b]{0.3\textwidth}
         \includegraphics[width=\textwidth]{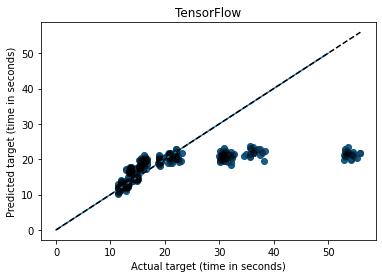}
         \caption{TensorFlow}
     \end{subfigure}
     \hfill
     \begin{subfigure}[b]{0.3\textwidth}
         \includegraphics[width=\textwidth]{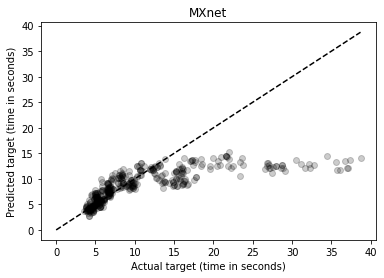}
         \caption{MXnet}
     \end{subfigure}
    \hfill
     \begin{subfigure}[b]{0.3\textwidth}
         \includegraphics[width=\textwidth]{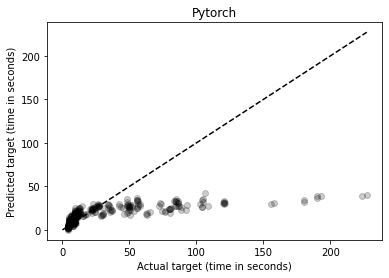}
         \caption{Pytorch}
     \end{subfigure}
     
\caption{Support vector machine prediction and the actual measured times in three deep learning frameworks \label{fig:6}}
    
\end{figure*}

\begin{figure*}[h!]
     \begin{subfigure}[b]{0.42\textwidth}
         \includegraphics[width=\textwidth]{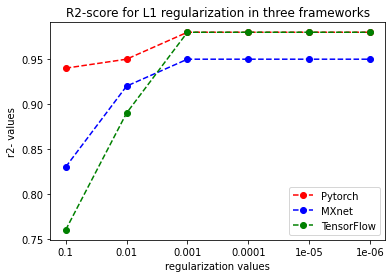}
         \caption{R2 values with different regularization values in three different frameworks using L1 regularization}
     \end{subfigure}
     \hfill
     \begin{subfigure}[b]{0.42\textwidth}
         \includegraphics[width=\textwidth]{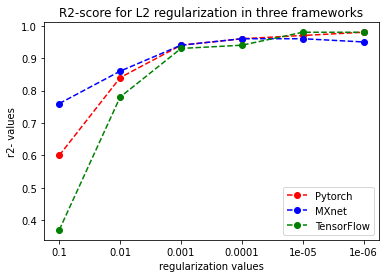}
         \caption{R2 values with different regularization values in three different frameworks using L2 regularization.}
     \end{subfigure}

\caption{Effect of regularization. \label{fig:7}}
    
\end{figure*}

\begin{figure*}[h!]
     \begin{subfigure}[b]{0.42\textwidth}
         \includegraphics[width=\textwidth]{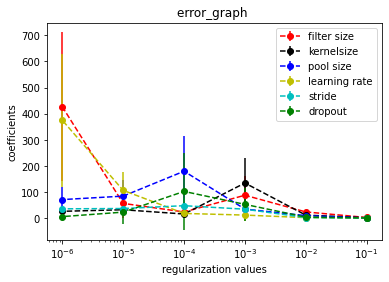}
         \caption{TensorFlow}
     \end{subfigure}
     \hfill
     \begin{subfigure}[b]{0.42\textwidth}
         \includegraphics[width=\textwidth]{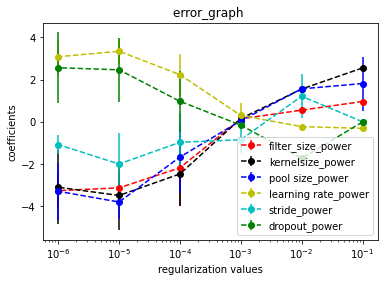}
         \caption{MXnet}
     \end{subfigure}
    \hfill
     \begin{subfigure}[b]{0.42\textwidth}
         \includegraphics[width=\textwidth]{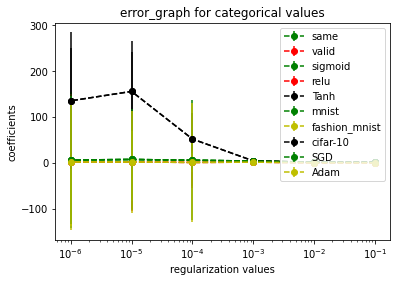}
         \caption{Pytorch}
     \end{subfigure}
\hfill
     \begin{subfigure}[b]{0.42\textwidth}
         \includegraphics[width=\textwidth]{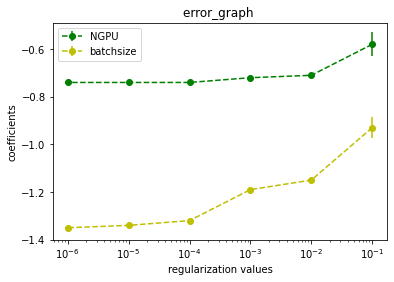}
         \caption{Pytorch}
     \end{subfigure}
     
\caption{Effect of regularization, with model coefficients plotted against regularization parameter. Constant coefficients of intrinsic parameters are plotted in (a), the power coefficients of intrinsic parameters are shown in (b), coefficients of categorical intrinsic parameters in (c), with powers of extrinsic parameters in (d). \label{Fig:8}}
    
\end{figure*}

\begin{table}[]
\caption{L1 and L2 regularization results in terms of the mean absolute percentage error (MAPE), Mean Squared Error (MSE), Root Mean Squared Error (RMSE)}\label{Table-IV}
\begin{tabular}{|l|lll|lll|}
\hline
     & \multicolumn{3}{l|}{L1 Regularization}                                  & \multicolumn{3}{l|}{L2 Regularization}                                  \\ \hline
     & \multicolumn{1}{l|}{MXnet}  & \multicolumn{1}{l|}{PYtorch} & TF & \multicolumn{1}{l|}{MXnet}  & \multicolumn{1}{l|}{PYtorch} & TF \\ \hline
MAPE & \multicolumn{1}{l|}{8\%}    & \multicolumn{1}{l|}{29\%}    & 13\%       & \multicolumn{1}{l|}{7\%}    & \multicolumn{1}{l|}{27\%}    & 10\%       \\ \hline
MSE  & \multicolumn{1}{l|}{105.74} & \multicolumn{1}{l|}{450.93}  & 220.16     & \multicolumn{1}{l|}{103.37} & \multicolumn{1}{l|}{443.35}  & 201.81     \\ \hline
RMSE & \multicolumn{1}{l|}{10.28}  & \multicolumn{1}{l|}{17.52}   & 14.83      & \multicolumn{1}{l|}{10.16}  & \multicolumn{1}{l|}{17.02}   & 14.20      \\ \hline
\end{tabular}
\end{table}

\subsection{Results and Analysis}
This section shows the results of our proposed performance model for three popular deep neural networks, i.e., TensorFlow, MXnet, and PyTorch. We evaluate the performance model with and without regularization and compare it with standard black-box regression models such as Support Vector Machine and Random Forest Regressor (RF). Tables~\ref{Table-II} and \ref{Table-III} compare intrinsic parameters and scalability in various frameworks with and without using regularization. Table~\ref{Table-IV} shows mean absolute percentage error values on predictions of the performance models using L1 and L2 regularization.


\subsubsection{Performance Evaluation of Deep Learning Frameworks using the Proposed Performance Model without regularization}
We have applied the differential evolution algorithm to our proposed model and evaluated it using the three deep learning frameworks. The actual execution time for training the model using the three frameworks is recorded, and predicted execution times are also generated. Fig.\ref{Fig:3} shows the scatter graph of the predicted execution times from the proposed model plotted against the actual execution time for the test dataset. The linear fit to the straight line determines how well the model can predict unseen configurations. We find the best fit constant coefficients for all frameworks are shown in Table \ref{Table-II}. 

The results show stable and consistent fits for the extrinsic parameters and the additive constant $C$, indicating that the scalability results are accurate. The higher variances in the intrinsic parameters are reduced by using regularization. Table \ref{Table-II} shows that the model gives broadly consistent performance for the constant coefficients, representing the relative importance of the process controlled by categorical parameters. For instance, Adam has a large constant for the activation function coefficients and takes more training time than SGD in Pytorch and TensorFlow frameworks, while SGD has the highest training time with the MXnet framework. The {\em padding} parameter, which is categorical with two possible values {\em valid} and {\em same}. {\em same} shows better performance for the {\em valid} mode. 

\subsubsection{Performance Evaluation of Deep Learning Frameworks using the Proposed Performance Model with regularization}
 We have applied regularization to the cost function to the proposed performance model to optimize the vector constants and reduce high variance in intrinsic parameters in three deep learning frameworks. We applied both regularizations to our model and compared the results of L1 and L2. The MAPE, MSE, and RMSE results are better with L2 regularization, as shown in Table \ref{Table-IV}. We therefore consider L2 regularization appropriate for our performance model and applied various regularization parameter values in logarithmic scale in L1 and L2 to find the best value of the $\lambda$ parameter. In Figures \ref{fig:7}(a) and \ref{fig:7}(b), we see that the R2 score deteriorates when the $\lambda$ value is higher than 0.001. For instance, when $\lambda$ = 0.001, the model fits well, and the model gives broadly consistent performance for the constant coefficients and represents the relative importance of the process controlled by categorical parameters. Furthermore, in Table \ref{Table-III}, we can see that the model gives consistent performance for the constant coefficients, representing the relative importance of the processes controlled by categorical parameters. The results show that the performance model using regularization is a generalised model with optimized good fits in all the frameworks. For example, for {\em padding} coefficients, {\em same} parameter takes more training time than {\em valid} parameter in all frameworks. For activation function coefficients, {\em Tanh} takes more training time than {\em Relu} and {\em Sigmoid} in MXnet and TensorFlow frameworks, while Relu takes maximum time with Pytorch. Also, in terms of dataset coefficients, the Fashion-MNIST dataset takes more training time than MNIST and CIFAR-10 datasets in all three frameworks. 
 
\subsubsection{Comparison of the Proposed Performance Model with Black Box Models}
We have compared the proposed model with two standard black box models, i.e., Random Forest Regressor and Support Vector Machine. Generally, the random forest regressor has better prediction accuracy due to its ensemble learning shown in Fig.\ref{fig:5}. The result shows a good linear fit compared to the differential evolution algorithm with and without regularization. However, the drawback of the random forest regressor is that it cannot give any insights into its internal working mechanism. Support vector machine regression is a non-parametric technique because it depends on kernel functionality. It is more productive in high-dimensional spaces. Fig.\ref{fig:6} shows the predicted and measured times of the support vector machine. The result shows a poor fit for all the deep learning frameworks compared with the random forest regressor and differential evolution algorithm with and without regularization. We evaluate the fits using the mean absolute percentage error between predicted execution time and actual times, as shown in Table \ref{Table-V}. Note that the performance of our proposed model is slightly inferior to the random forest. However, the proposed model can provide insights into the internal behaviour and scalability, which are impossible with a black box model such as a random forest.

\begin{table}[h!]
\centering
\caption{ Mean Absolute Percentage Error (MAPE) on predictions of the performance models on the 300 instances in the evaluation dataset in seconds.}\label{Table-V}
\begin{tabular}{|l|l|l|l|}
\hline
                        & \bf TensorFlow & \bf MXnet & \bf Pytorch \\ \hline
\begin{tabular}[c]{@{}l@{}}Differential\\ evolution\end{tabular} & 5\%  & 5\% & 12\%    \\ \hline
\begin{tabular}[c]{@{}l@{}}Differential\\ evolution\\ with regularization\end{tabular} & 10\%      & 7\%  & 14\%   \\ \hline

\begin{tabular}[c]{@{}l@{}}Random \\ forest\end{tabular}         & 0.7\%  & 3\%     & 23\%    \\ \hline
\begin{tabular}[c]{@{}l@{}}Support\\ vector\\machine\end{tabular}&16\%    &21\%  & 54\%   \\ \hline
\end{tabular}
\end{table}

\begin{table}[h!]
\centering
\caption{nGPUs scaling power in various frameworks, nGPUs represent number of GPUs.}\label{Table-VI}
\begin{tabular}{|l|l|l|}
\hline
\bf Frameworks & \bf nGPUs scaling power \\ \hline
TensorFlow & -0.74    \\ \hline
MXnet      & -0.99    \\ \hline
Pytorch    & -1.02    \\ \hline
\end{tabular}
\end{table} 
\subsubsection{Scalability Analysis}
Observing the coefficients $q$ from Table \ref{Table-II} and Table \ref{Table-III}, where $q$ is power in a multiplicative term representing an extrinsic parameter, we see that the extrinsic parameter coefficients are consistent in the proposed performance model with and without regularization. As shown in Table \ref{Table-VI}, -1 indicates ideal scaling, in which case the time is inversely proportional to the number of GPUs. The coefficients in Pytorch and MXnet frameworks show better scaling performance than TensorFlow. In TensorFlow, the value -0.73 is less than -1, indicating sub-optimal scaling.

\section{Conclusion and Future Works}\label{con}
In this work, we have developed a generic performance model for deep learning applications in a distributed environment with a generic expression of the application execution time that considers the influence of both intrinsic and extrinsic factors. We also formulated the proposed model as a global optimization problem and solved it using regularization on a cost function and differential evolution algorithm to find the best-fit values of the constants in the generic expression. 
The proposed model has been evaluated on three popular deep learning frameworks: TensorFlow, MXnet, and Pytorch, and shown to provide accurate performance predictions and interpretability. Also, the experimental results show that MXnet and Pytorch exhibit better scalability performance than TensorFlow.
Furthermore, the proposed method with regularization has been found to optimize the vector constants and reduce high variance in intrinsic parameters. The model can be applied to any distributed deep learning framework without requiring any changes to the code and can offer insight into the factors affecting deep learning application performance and scalability. Future work may include evaluating the model's performance on various deep learning frameworks to assess its generalisation capability.

\bibliographystyle{IEEEtran}
\bibliography{mybibfile1}


\EOD

\end{document}